\pdfoutput=1

\documentclass[11pt]{article}

\usepackage{acl}
\usepackage{booktabs}

\usepackage{times}
\usepackage{latexsym}

\usepackage[T1]{fontenc}

\usepackage[utf8]{inputenc}
\usepackage{tipa}
\usepackage{newunicodechar}
\newunicodechar{ʊ}{\textupsilon}

\usepackage{microtype}
\usepackage{graphicx}
\usepackage{svg}

\usepackage{inconsolata}
\usepackage{layouts}
\usepackage{lipsum}  

\usepackage{colortbl}
\usepackage{xcolor}
\usepackage{soul}
\usepackage{float}
\usepackage{dblfloatfix}    
\usepackage{nicematrix}

\usepackage{booktabs}
\usepackage{multirow}
\usepackage{makecell}

\usepackage[switch]{lineno} 

\title{Predicting positive transfer for improved low-resource speech recognition using acoustic pseudo-tokens}

\author{
\textbf{Nay San\textsuperscript{1},
Georgios Paraskevopoulos\textsuperscript{2},
Aryaman Arora\textsuperscript{1},
Xiluo He\textsuperscript{1},}\\
\textbf{Prabhjot Kaur\textsuperscript{3},
Oliver Adams\textsuperscript{4},
Dan Jurafsky\textsuperscript{1}
}
\\
\textsuperscript{1}Stanford University;
\textsuperscript{2}Athena Research Center;
\textsuperscript{3}Wayne State University;
\textsuperscript{4}Atos zData
\\
\texttt{nay.san@stanford.edu}
}

\begin{document}

\maketitle
\begin{abstract}

While massively multilingual speech models like wav2vec 2.0 XLSR-128 can be directly fine-tuned for automatic speech recognition (ASR), downstream performance can still be relatively poor on languages that are under-represented in the pre-training data.~Continued pre-training on 70--200 hours of untranscribed speech in these languages can help --- but what about languages without that much recorded data?~For such cases, we show that supplementing the target language with data from a similar, higher-resource `donor' language can help.~For example, continued pretraining on only 10 hours of low-resource Punjabi supplemented with 60 hours of donor Hindi is almost as good as continued pretraining on 70 hours of Punjabi.~By contrast, sourcing data from less similar donors like Bengali does not improve ASR performance.~To inform donor language selection, we propose a novel similarity metric based on the sequence distribution of induced acoustic units: the Acoustic Token Distribution Similarity (ATDS).~Across a set of typologically different target languages (Punjabi, Galician, Iban, Setswana), we show that the ATDS between the target language and its candidate donors precisely predicts target language ASR performance.

\end{abstract}

\section{Introduction}

For developing automatic speech recognition (ASR), `low resource' languages are typically classified as such based on the availability of transcribed speech.~Untranscribed speech, texts, or reliable metadata about the language are often assumed to be easily obtainable.~This assumption may not hold true for under-described languages with little digital representation.~For such languages, we are interested in two questions:~1) does leveraging untranscribed speech from a similar, higher-resource `donor' language for pre-trained model adaptation help improve speech recognition in the target language,~and 2) how do we select the best donor?

These questions are of interest as ASR system development with little transcribed speech has become viable with multilingual pre-trained transformer models for speech ~\cite[e.g. wav2vec 2.0 XLSR-128:][]{babu22_interspeech}.~Yet, as most languages are under-represented in the pre-training data, directly fine-tuning these models for ASR in the target language can yield lower performance compared to their well-represented counterparts \cite{conneau2023fleurs}.~While recent studies have shown the effectiveness of continued pre-training to adapt these models to the target language~\cite{NOWAKOWSKI2023103148,10301554}, they involved using 70--200 hours of target language data.~For some languages, it may be quite difficult to source this much speech data --- even untranscribed.

Thus, in our first set of experiments, we investigated whether supplementing target language data with data from another language could be a viable approach for model adaptation via continued pre-training (CPT).~We selected Punjabi as our target language to establish top-line performance when sufficient data \textit{is} available~\citep[70 hours, approximating the setup in][]{10301554}, along with a limited data baseline (when only 10 hours of Punjabi is available).~We compared this baseline to supplementing the 10 hours of Punjabi with 60 hours of data from 8 other Indic languages (Indo-Aryan:~Hindi, Urdu, Gujarati, Marathi, Bengali, Odia; Dravidian:~Malayalam, Tamil).~We fine-tuned each CPT-adapted model using the same 1 hour of transcribed Punjabi speech.

Results indicated that adding data from unrelated Dravidian languages (Malayalam, Tamil) or dissimilar Indo-Aryan languages (Bengali, Odia) yielded no better than baseline performance, 25\% word error rate (WER).~By contrast, we observed improved WERs from adding more similar languages (Marathi, Urdu, Gujarati, Hindi),~with adding Hindi coming close to the 70-hour Punjabi top-line: 23.2\% vs. 22.2\%, respectively.

In our second set of experiments, we investigated how well measures of similarity between the target and donor languages predicted target language ASR performance.~We found that commonly used measures based on external typological databases such as lang2vec \cite{littell2017uriel} were not sufficiently fine-grained for our use case and, crucially, also varied with the quality/completeness of the available metadata for a given target language.

To sidestep these issues, we propose the \textbf{Acoustic Token Distribution Similarity} (ATDS), which measures the degree of similarity for two untranscribed speech corpora based on frequencies of occurrence of recurring acoustic-phonetic sequences.~This measure extends Token Distribution Similarity~\cite{gogoulou2023study}, shown to correlate with positive transfer in continued pre-training for text-based language models.~To account for the text-/token-less nature of untranscribed speech corpora, we induce them in a bottom-up manner using wav2seq \cite{10096988}, a method for inducing pseudo-tokens using pre-trained speech embeddings.~We compared the ASR performance from the Indic language experiments to various similarity measures and found that ATDS offered the most accurate ranking.~Furthermore, ATDS correctly predicted the best donor language between two options for three non-Indic low-resource languages (Galician, Iban, and Setswana).\footnote{Appendix \ref{app:langs} provides an overview of the languages.}

In sum, the main contributions of this paper are:~1) a systematic study of pairwise transfer between languages in continued pre-training and its effects on target language ASR performance,~and 2) the development, analysis, and first validation of ATDS --- a fine-grained, bottom-up measure of acoustic-phonetic similarity to predict this ASR performance.~To facilitate reproducibility and further research,~we make available all our code, model checkpoints, and experimental artefacts.\footnote{\url{https://github.com/fauxneticien/w2v2-cpt-transfer}}

\section{Background: wav2vec 2.0}
\label{sec:background}

In this section, we provide a high-level overview of the wav2vec 2.0 model and highlight specific details about its architecture and training objectives that will be relevant to our later discussions.~Developed by \citet{baevski2020wav2vec}, wav2vec 2.0 is a type of \textit{self-supervised pre-trained transformer model}.~In machine learning, supervised learning involves the use of human-generated labels (e.g.~transcriptions), which can be time- and cost-intensive to create.~In self-supervised pre-training, the goal is to first train the model on a proxy task for which it can derive its own labels, before the resulting model is adapted or `fine-tuned' to the target task (e.g.~ASR).~Leveraging self-supervised pre-training has shown remarkable success across a variety of tasks when combined with a transformer-based model \cite{vaswani2017attention}, which excels at learning how various units in a sequence co-occur (e.g. words in a sentence).~Naturally, this has spurred on much experimentation for leveraging such models for low-resource ASR \cite[e.g.][]{coto-solano-etal-2022-development,guillaume2022fine,macaire-etal-2022-automatic,bartelds-etal-2023-making}.

As illustrated below in Figure~\ref{fig:w2v2}, the wav2vec 2.0 architecture consists of three parts:~1) a convolutional feature extractor that extracts learnable features from strided frames in the audio input,~2) a quantiser which clusters the audio features into into a set of discrete code vectors ($q_1$, ..., $q_5$),~and 3) a transformer attention network that learns context-enriched representations ($h_1$, ..., $h_5$).~For self-supervised pre-training, the model is optimised using a joint contrastive loss and a diversity loss.~The contrastive loss requires the model to use the neighbouring context to distinguish each masked frame amongst a set of negative distractors.\footnote{A speech variant of a Cloze test: e.g. given ``\textit{The big \_\_\_ chased the small rat.}'', select the correct answer from \{\textit{cat}, \textit{rat}, \textit{small}\}.}~The diversity loss requires the model to make equal use of all code vectors, preventing the model from relying on a small subset which can lead to trivial/sub-optimal solutions.

\begin{figure}[!hb]
    \includegraphics[width=1.0\linewidth]{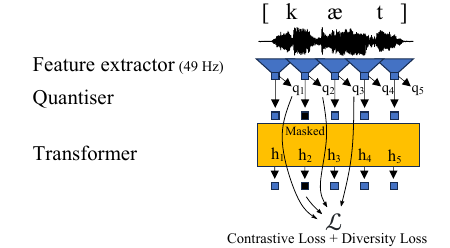}
    \caption{Illustration of the wav2vec 2.0 architecture. Adapted from \newcite{baevski2020wav2vec}.}
    \vspace{-0.5em}
    \label{fig:w2v2}
\end{figure}

We highlight here two important details relevant for our later discussions on acoustic tokens.~The first is that the representations learned by the transformer network are particularly useful for speech applications requiring fine-grained comparisons of acoustic-phonetic content,~e.g.~speech information retrieval \cite{san2021leveraging}, second language pronunciation scoring \cite{bartelds2022neural,richter2023relative}, spoken dialect classification \cite{bartelds-wieling-2022-quantifying,guillaume23_sigul} --- particularly those learned by the middle layers of the transformer network (e.g.~layers 12--16 of a 24-layer network).

The second detail is that these representations are encoded as vectors in a high-dimensional latent space and emitted at a rate of 49 Hz.~Combined with the diversity loss that requires exploration of this space, there is a many-to-many relationship between phonetic categories and these vectors --- both in the latent space and in time.~For example, as illustrated in Figure \ref{fig:w2v2}, a single speech sound such as [\ae] may last for three time steps (yielding $h_2, h_3, h_4$).~The first two vectors ($h_2, h_3$) may be very close in the latent space, as they map to the start of [\ae] sounds while the third $h_4$ may map to [\ae] sounds preceding [t] and is thus located in a different part of the latent space.~Thus to derive phone-like tokens from these representations, we must group them in the latent space (e.g.~using \textit{k}-means clustering) and then again in time according to how the grouped units themselves routinely co-occur (e.g.~using subword modelling).

~In this way, the goal of tokenisation for both text and speech is driven by the need to derive units of a practical granularity based on the nature of the input: from coarser-grained words to finer-grained sub-words in text and, inversely, finer-grained sub-phones to coarser-grained phones in speech.~We return to these two details in our development of the Acoustic Token Distribution Similarity measure for comparing the acoustic-phonetic similarity of two untranscribed speech corpora.

\section{A systematic study of pairwise transfer}

\subsection{Motivation}
\label{sub_sec:pair_motivation}

Since the release of the original English wav2vec 2.0 model pre-trained on the 960 hour LibriSpeech corpus~\cite{7178964}, additional massively multi-lingual variants have also been developed: XLSR-53, pre-trained on 56k hours from 53 languages~\cite{conneau21_interspeech}; XLSR-128, pre-trained on 436k hours from 128 languages~\cite{babu22_interspeech}; and MMS, pre-trained on 491k hours from 1,406 languages \cite{pratap2023scaling}.~In each case, the vast majority of the pre-training data is drawn from European language sources.

Given the under-represented nature of most languages in these wav2vec 2.0 models, several studies have investigated continued pre-training (CPT) to adapt them for target languages \cite[e.g.][]{javed2021building,dehaven2022improving,NOWAKOWSKI2023103148,bartelds-etal-2023-making,10301554}.~\citet{NOWAKOWSKI2023103148} adapted the XLSR-53 model using 200 hours of Ainu.~Using the same 40 minutes of transcribed Ainu for ASR fine-tuning,~they found that the adapted model resulted in a 8.8\% absolute word error rate (WER) decrease over the off-the-shelf XLSR-53 model.~For many low resource languages, however, obtaining 200 hours of speech data may not be feasible.

In their study of unsupervised domain adaptation for Greek,~\citet{10301554} found adapting the XLSR-53 model via CPT using a small 12-hour dataset of Greek read speech to be ineffective.~However, they found that successful CPT-based adaptation could be achieved with the use of multi-domain data,~e.g.~12 hours of read speech mixed with 70 hours of newscasts.~Given these findings, we were motivated to investigate whether comparable results could be achieved by supplementing target language data with data from another language.

\subsection{Method}

\subsubsection{Model training}

As our primary interest was in examining how downstream ASR performance is affected by the dataset(s) used in continued pre-training (CPT), we carried out nearly identical experiments as those in \citet{NOWAKOWSKI2023103148} by obtaining their configuration file for wav2vec 2.0 pre-training using the official fairseq library.\footnote{\url{https://github.com/facebookresearch/fairseq}}~Similarly, we follow the official fine-tuning recipe suitable for 1 hour of transcriptions.~For both procedures, we made modifications to suit our hardware configuration and compute budget, as detailed in Appendix \ref{app:methods}.

\subsubsection{Data}

For a systematic investigation of how downstream ASR performance in a given target language varies with the choice of donor language added during CPT, we required a dataset with some specific characteristics: a) contains a variety of both similar and dissimilar languages with, b) at least 60 hours per language, c) and collected in relatively similar acoustic conditions, and d) not have already been used in the original pre-training process.~Accordingly, for this set of experiments, we sourced data from IndicSUPERB~\cite{10.1609/aaai.v37i11.26521},
a dataset of 12 Indic languages containing 65--180 hours of read speech per language (Indo-Aryan:~Bengali, Gujarati, Hindi, Marathi, Odia, Punjabi, Sanskrit, Urdu;~Dravidian:~Kannada, Malayalam, Tamil, Telugu).

Amongst these 12 IndicSUPERB languages, we selected Punjabi as our target language as it is relatively low-resourced (cf.~Hindi, Tamil), its geographical and typological location (yielding a variety of closer and farther geographic/typological distances to the others), and also native speaker-linguist expertise on our research team for data validation and error analysis.

As illustrated below in Figure \ref{fig:indic_data}, we began by selecting for our top-line condition a random 70h subset of Punjabi from the total 136h available in IndicSUPERB, and then from this subset a random 10h selection for our baseline, and again a random 1h subset for fine-tuning.~We also selected a random 1h validation and 2h test set both disjoint from each other and any data to be used for pre-training.

\begin{figure}[H]
    \centering
    \includegraphics[width=1\linewidth]{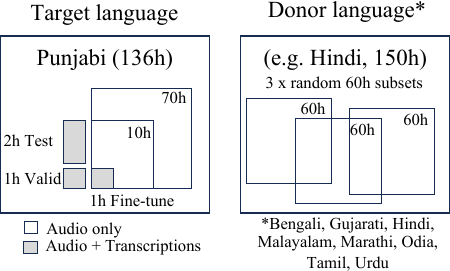}
    \caption{Data selection for transfer experiments}
    \label{fig:indic_data}
\end{figure}

Additionally, for each donor language (Bengali, Gujarati, Hindi, Malayalam, Marathi, Odia, Tamil, Urdu), we created three random 60h subsets.~Given the total amounts of data available in IndicSUPERB for each language (e.g.~87h for Urdu but 129h for Gujarati), there is however some unavoidable overlap between these subsets for each language (i.e.~they are not disjoint 60h splits).~Using each of the 60h subsets, we conducted 3 separate CPT runs per language to obtain estimates for both within- and between-donor language differences on downstream ASR performance.

\newpage

\subsection{Results and discussion}

\begin{figure*}[!ht]
\definecolor{mygray}{gray}{0.9}
\centering
\vspace{1em}
\scalebox{0.95}{
\renewcommand{\arraystretch}{0.95}
\begin{tabular}{@{}llcl@{}}
\toprule
\multirow{1}{*}{\makecell{\vspace{-0.45em}\\Condition}}     & \multicolumn{2}{c}{Test set WER (WERR)}                                                & \multirow{1}{*}{\makecell{\vspace{-0.45em}\\Data for continued pre-training}}          \\ \cmidrule(r){2-3}  
                               & \multicolumn{1}{c}{Median}         & \multicolumn{1}{c}{Range}                     &                         \\ \midrule

T.\hspace{0.87em}In-domain top-line         & 22.2 (11.2\%)  & -                                                                      & 70h Punjabi                                    \\  \cmidrule{1-4} 
E1. Most similar   & \textbf{23.5 (6.0\%)}     & 23.4--23.8   & 10h Punjabi + 60h Hindi                          \\ \cmidrule{1-4}
\multirow{3}{*}{\makecell{\vspace{-0.45em}\\E2. Similar}}   & 24.4 (2.4\%)   & 24.3--24.5     & 10h Punjabi + 60h Urdu                            \\ \cmidrule{2-4}
   & 24.4 (2.4\%)   & 24.2--24.4   & 10h Punjabi + 60h Gujarati                       \\  \cmidrule{2-4}
   & 24.6 (1.6\%)   & 24.5--24.7 & 10h Punjabi + 60h Marathi                         \\  \cmidrule{1-4}

\rowcolor{mygray} B.\hspace{0.75em}Only target data baseline & 25.0     & -                                                                      & 10h Punjabi                                      \\  \cmidrule{1-4} 
\multirow{4}{*}{\makecell{\vspace{-0.1em}\\E3. Unrelated/dissimilar}} & 25.0 (0.0\%)     & 25.0--25.2    & 10h Punjabi + 60h Odia                             \\  \cmidrule{2-4} 
         & 25.1 (-0.4\%)  & 25.0--25.4    & 10h Punjabi + 60h Tamil                            \\ \cmidrule{2-4}
         & 25.1 (-0.4\%)  & 25.0--25.3    & 10h Punjabi + 60h Malayalam                        \\  \cmidrule{2-4}
& 25.2 (-0.8\%)  & 25.1--25.2  & 10h Punjabi + 60h Bengali    \\ 
 \cmidrule{1-4}
U.\hspace{0.75em}Unadapted XLSR-128     & 30.8 (-23.2\%) &  -                                                                     & -                                               \\ \bottomrule
\end{tabular}
}

\captionof{table}{Automatic speech recognition (ASR) results from fine-tuning wav2vec 2.0 XLSR-128 \cite{babu22_interspeech} with and without adaptation via continued pre-training (CPT). CPT-adapted models were trained for 10k updates using 70 hours of Punjabi for the topline (T), 10 hours of Punjabi for the baseline (B), and 10 hours of Punjabi combined with 60 hours of data from another language for the experiment conditions (E1, E2, E3).~All models were fine-tuned with the same 1 hour of Punjabi data.~ASR performance reported in word error rate (WER) and relative word error rates (WERR), relative to the 10 hour CPT baseline (B).~For each experiment condition, median and range were obtained from 3 CPT runs per language with different donor data in each run.}

\label{tab:indic_results}

\end{figure*}

Compared to directly fine-tuning the XLSR-128 model, adapting the model first via continued pre-training (CPT) with 70 hours of Punjabi yields a large improvement in downstream ASR performance (unadapted 30.8\%~vs.~22.2\% CPT-adapted, 70h).~This constitutes an absolute WER difference of 8.6\% and is consistent with improvements reported in previous CPT experiments \cite[][]{NOWAKOWSKI2023103148,10301554}.~We found that model adaptation with only 10 hours of Punjabi still yielded an appreciable improvement over the unadapted model:~5.8\% absolute (unadapted 30.8\%~vs.~25.0\% CPT-adapted, 10h).

We now turn to our experiment conditions in which 10h of Punjabi is supplemented with 60h of data from another language.~As summarised below in Table \ref{tab:indic_results}, the effects of donor language on Punjabi ASR performance can be divided into roughly three strata.~In the bottom stratum (E3), we find unrelated Dravidian languages (Tamil, Malayalam) or dissimilar Indo-Aryan languages (Bengali, Odia).~When data from these languages are added during CPT, we find no meaningful difference compared to using only 10 hours of Punjabi (WERR: -0.8--0.0\%).

In the middle stratum (E2), we find relatively similar Indo-Aryan languages (Marathi, Gujarati, Urdu).~When data from these languages are added, we find modest improvements over using only 10 hours of Punjabi (WERR: 1.6--2.4\%).~In the top stratum (E1), we find Hindi --- the most similar Indo-Aryan language amongst our candidate donors.~When data from Hindi is added, we find a large improvement over using only 10 hours of Punjabi (WERR: 6.0\%).~In fact, adding the 60h of Hindi results in ASR performance that is in absolute terms close to the 70h Punjabi topline:~23.5\%~vs.~22.2\%,~respectively.

We have established that there are observable differences in target language ASR performance that vary with the donor language added during continued pre-training and that these differences appear to align with qualitative notions of language similarity.~In the next section, we investigate quantitative measures of similarity and evaluate their correlations to these observed differences.

\section{A bottom-up approach to similarity}

\subsection{Motivation}

A common method for calculating similarities between languages is to use language vectors from the lang2vec database \cite{littell2017uriel}, which itself draws on other databases \cite[e.g.~phonological information from WALS:][]{haspelmath2009typological}.~\citet{9688276} investigated how well measures based on lang2vec and other data sources correlated with successful transfer learning for ASR.~Of the lang2vec similarity metrics, they found that genetic and geographic measures correlated highly with better ASR performance but, surprisingly, inventory and phonological measures did not.~Acoustic similarities as derived from embeddings of a pre-trained spoken language identification model were also found to correlate strongly with better ASR performance.~In the context of continued pre-training, we questioned whether the to-be-adapted model could be used for this purpose.

\begin{figure*}[!ht]
\centering

\includegraphics[width=1\textwidth,height=3in]{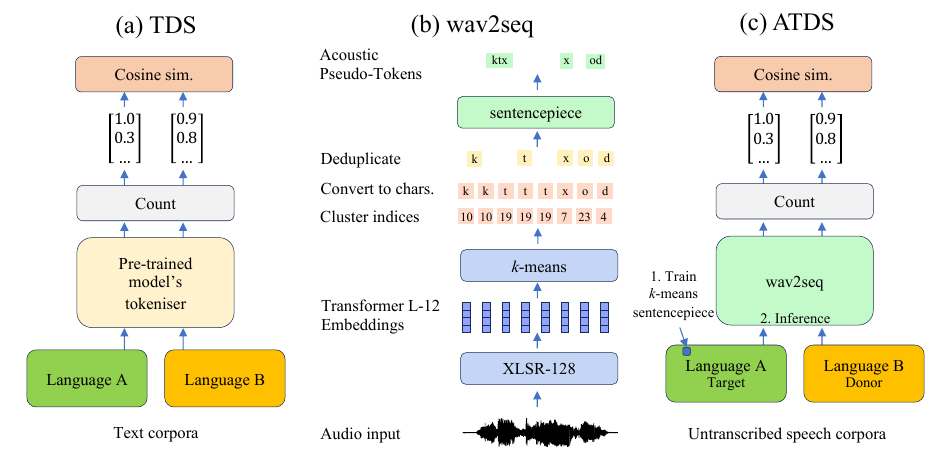}
\caption{Derivation of the Acoustic Token Distribution Similarity (ATDS) measure for predicting positive transfer between two languages resulting from continued pre-training (CPT) of a pre-trained speech model (e.g. XLSR-128).~ATDS extends to the speech domain the concept of Token Distribution Similarity \cite[TDS:][]{gogoulou2023study}, shown to predict positive transfer for CPT in the text domain.~To account for the token-less nature of untranscribed speech, pseudo-tokens are derived using the wav2seq process proposed by \citet{10096988}.~As is the case for text tokenisation, the goal is to derive units of practical granularity based on the raw input.~While typical text tokenisation sub-divides words into sub-words (e.g.~\emph{eating} $\rightarrow$ \emph{eat,ing}), the analogous process for raw speech data involves grouping sub-frames into more phone-like units: first based on featural similarity (e.g.~using a k-means model on embeddings) then again based on distributional similarity (e.g.~using a sub-word modelling).}
\label{fig:atds}

\end{figure*}

Investigating an analogous question in the text domain, \citet{gogoulou2023study} evaluated various measures for predicting transfer characteristics for transformer language models initially pre-trained on one language (e.g. English) and subsequently adapted to another (e.g. Icelandic), and how these characteristics varied according to the distributions of data in the respective language corpora.~They propose a novel metric:~the Token Distribution Similarity (TDS), which correlated with positive transfer.~As illustrated below in Figure \ref{fig:atds} (a), the TDS is derived by 1) using the pre-trained model's tokeniser to process a sample of data from each language, 2) then generating a token frequency vector for each language, and 3) taking the cosine similarity between these two vectors.~Given these promising results for predicting positive cross-lingual transfer for continued pre-training on text, we investigated whether they extended to the speech domain.

\vspace{0.25em}
\subsection{Induction and analysis of acoustic tokens}
\vspace{0.25em}

In order to compute token distribution similarity for two untranscribed speech corpora, we first need to `tokenise' the corpora.~For this purpose, we can leverage speech representations extracted using a middle transformer layer (e.g. Layer 12 of 24) of a pre-trained wav2vec 2.0 model such as XLSR-128.~As previously highlighted above (in §\ref{sec:background}), these representations are useful for fine-grained comparisons of acoustic-phonetic content and, to make use for these representations for inducing phone-like tokens, they must first be grouped in the high-dimensional latent space and then again in time based on their co-occurrences.

~Accordingly, we use the wav2seq procedure proposed by \citet{10096988}.\footnote{Originally developed for inducing pseudo-tokens for use in a self-supervised pseudo speech recognition task for jointly pre-training a speech encoder and text decoder.}~As illustrated below in Figure \ref{fig:atds} (b), the first step is using a pre-trained model (e.g.~XLSR-128) to extract speech embeddings from a transformer layer.~The second step involves training a \textit{k}-means model to cluster these embeddings (i.e.~group similar sounds together).~The cluster indices are converted to characters (by a simple Unicode table lookup,~e.g.10$ \rightarrow $\textit{k}, 19$\rightarrow$\textit{t}, etc.)~and then deduplicated.~To discover frequently occurring sound sequences, the third step involves using these character strings to train a subword model \cite[e.g.~via sentencepiece:][]{kudo2018sentencepiece}.~Once these models are trained on a subset of the corpus, they are used to derive pseudo-tokens for the rest of the corpus and, in our application, also for candidate donor corpora.

To discover whether these pseudo-tokens exhibited both within- and cross-language consistency, we conducted an analysis using the Punjabi and Hindi datasets of the CommonVoice corpus \cite{ardila2019common}, for which forced-aligned phoneme labels are available in the VoxCommunis corpus \cite{ahn-chodroff-2022-voxcommunis}.~For this analysis, we trained the \textit{k}-means and subword models on Punjabi (the target language) and then induced pseudo-tokens for both Punjabi and Hindi.\footnote{Appendix \ref{app:methods} details the analysis procedure.}

Results of our analyses revealed that the most frequent pseudo-token in Punjabi, $t_1$, consistently corresponded to the /\textscripta/ label,~i.e.~$P(\textrm{/\textscripta/}|t_1)$ = 0.69.~Similarly, we found that $t_2$ consistently corresponded to high-back vowels:~i.e.~$P(\textrm{/o/}|t_2)$ = 0.56,~$P(\textrm{/u/}|t_2)$ = 0.18,~$P(\textrm{/\textupsilon/}|t_2)$ = 0.09.~For Hindi, we found that $t_2$ also consistently corresponded to the same vowel labels, while $t_1$ consistently corresponded to /a\textlengthmark/ --- also a low vowel.~Such minor differences are likely attributable to VoxCommunis labels being automatically derived via grapheme-to-phoneme conversion and thus do not represent narrow phonetic transcriptions.~Given their broad categorical consistency, these tokens may be useful for cross-language comparisons.

\subsection{Acoustic Token Distribution Similarity}

\begin{figure*}[!ht]

\centering
\vspace{1em}
\includegraphics[width=1\textwidth]{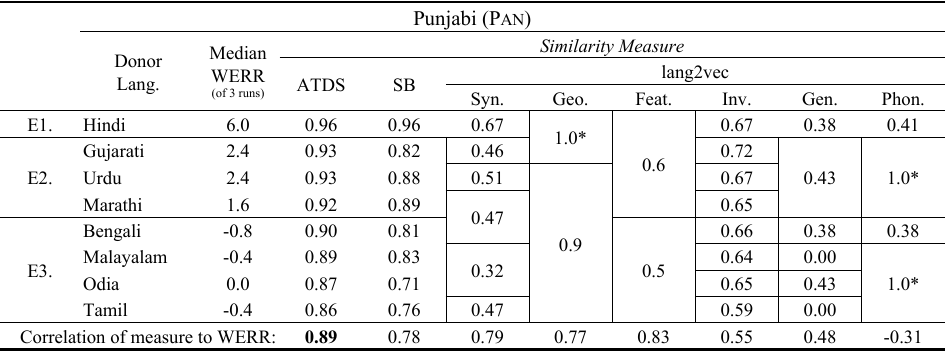}
\captionof{table}{Acoustic Token Distribution Similarity (ATDS) measure between Punjabi and donor language predicts downstream speech recognition performance as measured by relative word error rate (WERR) when fine-tuning the wav2vec 2.0 XLSR-128 model adapted using continued pre-training (CPT) on 10 hours of target and 60 hours of donor language speech.~Other similarity measures for comparison are derived embeddings of the SpeechBrain language identification model and from the lang2vec database (syntactic, geographic, featural, inventory, genetic, and phonological).~* indicate erroneous similarity scores resulting from identical, imputed vectors within the database.~Correlations (Pearson's \textit{r}) calculated using 24 data points (8 donor languages x 3 CPT runs per language with different donor data in each run).}
\label{tab:atds_results}

\end{figure*}

We propose the \textit{Acoustic} Token Distribution Similarity (ATDS) measure, which, as illustrated above in Figure \ref{fig:atds} (c), is a straightforward composition of wav2seq \cite{10096988} and Token Distribution Similarity \cite[TDS:][]{gogoulou2023study}.~We provide two analyses showing that the ATDS between a target language and its candidate donors precisely predicts downstream ASR performance in the target language resulting from continued pre-training of a speech model on target and donor data.

In our first analysis, we examined how well ATDS can account for the results of the Indic language experiments and how ATDS compares to other measures.~Accordingly, for ATDS, we trained the relevant wav2seq models on Punjabi (the target language), induced tokens on Punjabi and all donor languages, then calculated the token frequency vectors and computed the pairwise cosine similarities.~Similar to \citet{9688276}, we also computed similarities using lang2vec and corpus-level means of utterance-level embeddings extracted using a pre-trained spoken language identification model \cite[SpeechBrain:][]{speechbrain}.

Results of this analysis revealed that the two bottom-up acoustic measures provide overall a finer-grained ranking than top-down lang2vec measures.~For example, as shown above in Table \ref{tab:atds_results} Feat. (a combination of all lang2vec data sources), the featural similarity, splits the four languages into two groups:~similar~(0.6:~Hindi--Marathi) and dissimilar (0.5:~Bengali--Tamil).~Additionally, we also found erroneous, perfect similarity values (1.0), resulting from identical language vectors for the category (e.g. Phon.: Marathi-Punjabi), likely an artefact of data imputation.~These results demonstrate that while top-down data may be suitable for more noise-tolerant applications (e.g. large-scale typological comparisons), they may not be well suited for helping select donors for a specific under-described target language if the relevant metadata is unavailable or inaccurate.

While both acoustic measures accurately select Hindi as the most suitable donor, ATDS provides a better ranking of the donor languages.~As shown in Table \ref{tab:atds_results} (SB), the measure based on SpeechBrain embeddings ranks Gujarati as being as dissimilar to Punjabi as Bengali/Malayalam.~We make a similar observation as above that using embeddings from a different pre-trained model than the one to be adapted via CPT risks adding unwanted noise to an inherently hard task.~Leveraging the representations of the pre-trained model to be adapted reduces this risk, reflected in ASR improvement being most correlated with ATDS (\textit{r}~=~0.89).

\begin{figure*}[!ht]

\centering
\vspace{0.2em}
\includegraphics[width=1\textwidth]{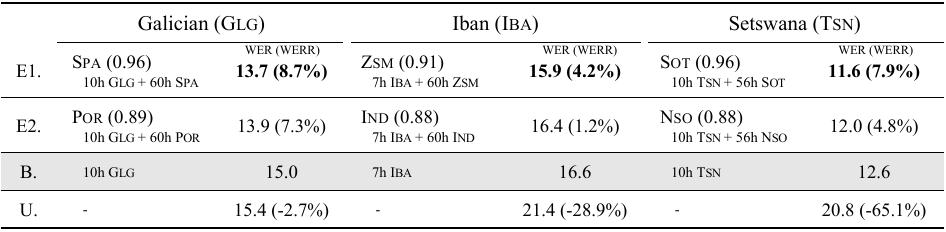}
\captionof{table}{Validation of the Acoustic Token Distribution Similarity (ATDS) measure for predicting target language automatic speech recognition (ASR) performance as a result of continued pre-training (CPT) of the wav2vec 2.0 XLSR-128 model using mix target and donor language data.~For each target language (Galician, Iban, Setswana), U. indicates ASR performance from using the unadapted XLSR-128 model, B. indicates performance from CPT adaptation with only target language data (7--10 hours), and E1--E2 using target language data supplemented with 56--60 hours of donor language data.~Parentheses next to donor language names indicate ATDS to the target language.~Percentages within parentheses indicate relative word error rate (WERR), relative the baseline word error rate (B) within the same column.~Donor language codes are: Spanish (\textsc{Spa}), Portuguese (\textsc{Por}), Malay (\textsc{Zsm}), Indonesian (\textsc{Ind}), Sesotho (\textsc{Sot}), Sepedi (\textsc{Nso}).}

\label{tab:valid_results2}

\end{figure*}

In our second analysis, we examined whether the ATDS measure generalised beyond the Indic languages through identical CPT experiments on a typologically varied set of target languages.~We selected triplets of languages consisting of 1) a target language,~2) a language more similar to the target as measured by ATDS,~and 3) another language relatively farther.~As summarised below in Table \ref{tab:valid_results2}, the target languages were Galician (West-Iberian), Iban (Malayic), and Setswana (Sotho–Tswana).~For Galician, ATDS predicted that Spanish (\textsc{Spa}: 0.96) was more similar than Portuguese (\textsc{Por}: 0.89); for Iban, Malay (\textsc{Zsm}: 0.91) more than Indonesian (\textsc{Ind}: 0.88); and for Setswana, Sesotho (\textsc{Sot}: 0.96) more than Sepedi (\textsc{Nso}: 0.88).

Results of these CPT experiments are summarised below in Table \ref{tab:valid_results2}.~We first note the large difference between Galician and the other languages in the improvement yielded by CPT baselines (B) compared to fine-tuning the unadapted XLSR-128 (U).~As we sourced Galician data from CommonVoice (on which XLSR-128 was already pre-trained), CPT yields little further gain (U. 15.4\% vs. B. 15.0\%).~By contrast, ASR performance was much improved via CPT adaptation for both Iban (U. 21.4\% vs. B. 16.6\%) and Setswana (U. 20.8\% vs. B. 12.6\%).~These results constitute further evidence that directly fine-tuning massively multilingual models can yield sub-optimal performance for under-represented languages and that continued pre-training can help close this performance gap.

We found that the ATDS predictions are borne out for all three target languages (even for Galician in spite of relatively reduced benefits).~As shown above in Table \ref{tab:valid_results2} for each of the target language columns (Galician, Iban, Setswana), larger improvements in target language ASR performance are observed as a result of continued pre-training on target language data supplemented with data from a more similar language as measured by ATDS than a less similar one (respectively, rows E1. vs. E2).~Combined with results for Punjabi above, our findings altogether provide strong evidence for the effectiveness of ATDS for predicting positive transfer between target and donor languages for CPT-based model adaptation.

\section{Limitations and future directions}

We limited the scope of this paper to exploring the transfer between languages and as such used the standard wav2vec 2.0 pre-training recipe for model adaptation.~We acknowledge that this requires a large compute budget beyond what is affordable in many low resource scenarios.~In future, we hope to investigate whether or to what extent transfer learning can be combined with more compute-efficient adaptation methods.

To conduct a systematic study of pairwise transfer, we used domain-matched, high-quality ASR datasets containing mostly read speech and only examined sourcing data from a single donor language.~Questions relating to multi-donor and multi-domain transfer and how such interactions affect downstream performance in the target language will need to be addressed in future research. 

\section{Conclusion}

For developing automatic speech recognition (ASR) systems for languages with very few resources, we demonstrated that massively multilingual pre-trained models for speech can be successfully adapted via continued pre-training using a mix of data from the target language and supplemental data from a similar, higher-resource `donor' language.~Additionally, we motivate and propose the Acoustic Token Distribution Similarity (ATDS) --- a novel measure of similarity to predict positive transfer between a donor and target language.~Across a set of typologically different target languages (Punjabi, Galician, Iban, Setswana), we show that the ATDS between the target language and its candidate donors precisely predicts target language ASR performance.

We attribute this predictive capability of ATDS to leveraging the knowledge of the pre-trained model to be adapted, its inductive biases and training objectives, and the distributions within the candidate datasets that will be used to adapt it.~It is indeed expected that this measure then is able to predict downstream task improvements better than measures based on other models or external information --- the latter of which is not always available or reliable for under-described languages.~Given the high cost associated with continued pre-training, however, we argue that using a well-calibrated, task-specific measure minimises the chance of costly, unexpected outcomes.

We make a final observation here that various target-donor language pairs where we observed successful transfer exist in linguistic situations with significant, sustained contact.~For example, Galician is a minoritised language in Spain and virtually all Galician speakers are bilingual in Spanish \cite{de2020phonetic}. Similarly, \citet{macaire-etal-2022-automatic} report that fine-tuning a model pre-trained on French was particularly successful for Gwadloupéyen and Morisien, two French-based creole languages.~These findings suggest that to develop truly inclusive speech technologies in a resource efficient manner, what will be required is a nuanced understanding of what factors linguistic and non-linguistic yield sufficiently high levels of cross-lingual similarity which in turn permit positive transfer.

\section*{Acknowledgements}

We thank Karol Nowakowski for helpful correspondence and also making available experiment configuration files. We thank Tol\'{u}l\d{o}p\d{\'{e}} \`{O}g\'{u}nr\d{\`{e}}m\'{i} for assistance with using the Stanford NLP cluster as well as Stanford UIT for providing Google Cloud research credits, both without which extensive continued pre-training experiments would not have been possible.~We are very grateful for support from both Alexis Michaud at the Centre National de la Recherche Scientifique (CNRS, France) and the Institut des langues rares (ILARA) at École Pratique des Hautes Études.~We are thankful for the continued support from Dr. Weisong Shi from University of Delaware.

\bibliography{anthology,custom}

\appendix

\section{Materials and methods}
\label{app:methods}

\subsection{Data}

Galician, Spanish, Portuguese, and Indonesian data were sourced from CommonVoice \cite{ardila2019common}; Setswana, Sesotho, and Sepedi from NCHLT \cite{barnard2014nchlt}; Malay from MASS \cite{tan2009mass}; and Iban from \citet{juan2015using}.~For Iban only 7 hours of target data was available and for Sesotho/Sepedi only 56 hours per language of donor data was available.~Experiments were otherwise identical to the Indic experiments.

\subsection{Continued pre-training}

We carried out nearly identical experiments as the single-language experiments in \citet{NOWAKOWSKI2023103148} for Ainu, as we obtained their configuration file for wav2vec 2.0 pre-training using the fairseq library.\footnote{\url{https://github.com/facebookresearch/fairseq}}~We made approriate modifications to suit our hardware configuration (4 x A100 40GB GPUs), setting the batch size to 1.5M samples per GPU and gradient accumulation to 16 steps, yielding an effective batch size of 100 minutes.

As in other wav2vec 2.0 multilingual pre-training configurations~\cite{conneau21_interspeech,babu22_interspeech,10.1609/aaai.v37i11.26521}, we form multi-lingual batches (specifically, \textit{bi-}lingual in our case).\footnote{Specifically, we use the implementation adapted from \url{https://github.com/AI4Bharat/IndicWav2Vec/}}~We set our sampling alpha to 0.0, which results in data being drawn uniformly from the two languages (i.e.~target is over-sampled).~We make this modification based on the CPT method in \citet{10301554}, where in- and out-of-domain Greek data were evenly sampled in each batch.~In this way, we consider this method akin to ``similar-language regularisation''~\cite{neubig-hu-2018-rapid}, as we are more concerned with preventing over-fitting rather than learning about the donor data.

For each CPT run, we start from the official XLSR-128 model checkpoint and update the model for 10k steps.~We determined this value based on our pilot runs.~We found that 10k updates were sufficient to observe improved downstream ASR performance comparable to previous results \cite[e.g.][]{NOWAKOWSKI2023103148}.~This choice permitted us to maximise the number of languages compared in this paper, as each run required on average 15 hours (for 10k steps).
For select runs, we also verified that no further improvements could be obtained with additional updates (up to 20k).

\subsection{ASR fine-tuning and evaluation}

For ASR fine-tuning, we follow the official wav2vec 2.0 fine-tuning recipe suitable for 1 hour of transcriptions, modified for our hardware setup.~We use a single A6000 48 GB GPU with a batch size of 5.6M samples per GPU and accumulate gradients for 2 steps, yielding an effective batch size of 11.6 minutes.~As standard, the feature extractor is kept frozen across all updates and the transformer is frozen for the first 10k of 13k total updates, optimised using a CTC loss.~Each fine-tuning run required on average 3.5 hours.~For our findings to be applicable for languages with little external text data, we use Viterbi decoding without a language model to obtain transcriptions from the fine-tuned model for evaluation.

\subsection{ATDS analyses}

For all analyses, we trained the necessary \textit{k}-means and sentencepiece models on a random 5-hour subset of target language data (except for CommonVoice Punjabi, which had in total 4 hours available in the latest 15.0 release).~We adopted hyperparameters based on previous findings: extracting embeddings from the mid-point layer (12 of 24) of the XLSR-128 model \cite[e.g.][]{san2021leveraging,bartelds2022neural}, and \textit{k}=500 for \textit{k}-means and V=10k for the subword model which were reported as optimal values in \citet{10096988}.~Using a 12GB 3060 GPU,~embedding extraction required about 10 minutes per data subset. The \textit{k}-means models trained in about 8 minutes and subword models in less than a minute.

For CommonVoice (CV) Punjabi and Hindi, we conducted similar analyses as those reported by \citet[][Appendix D]{baevski2020wav2vec} for analysing correspondences between the wav2vec 2.0 code vectors and hand-aligned phone labels from TIMIT \cite{Garofolo1993}.~In our case we used the labels for CV Punjabi and Hindi via grapheme-to-phoneme conversion, forced-aligned to the audio, and released as Praat TextGrids in the VoxCommunis corpus \cite{ahn-chodroff-2022-voxcommunis}.~We then added tiers containing the wav2seq-induced labels.~We hand-inspected several TextGrids for data validation then compiled the correspondences between the wav2seq induced tokens and phoneme labels.~We make available all TextGrids as well as the aggregated data.

\newpage

\section{Languages}
\label{app:langs}

\begin{figure*}[!t]
    \centering
    \includegraphics[width=\textwidth]{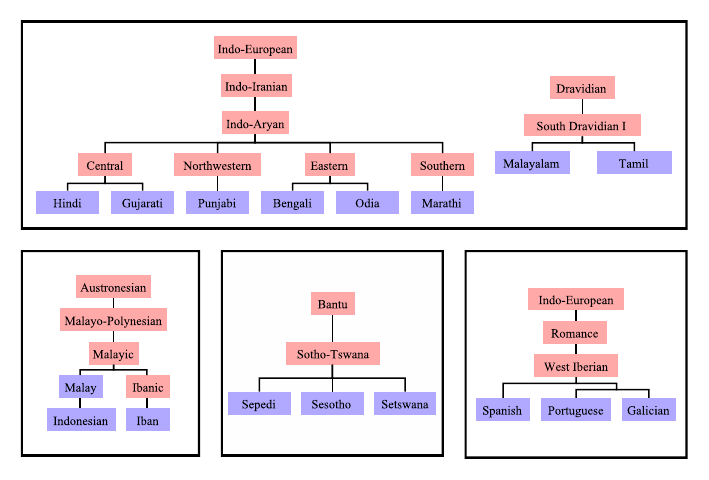}
    \caption{Family trees of the languages studied in this paper. Language families are in red and languages are in blue.}
    \label{fig:lang-trees}
\end{figure*}

\subsection{Indo-Aryan and Dravidian}

Punjabi (\textsc{Pan}) is a Northwestern Indo-Aryan language spoken by over 100 million people along the five major tributaries of the Indus river, spanning the state of Punjab in India and the province of Punjab in Pakistan.~Hindi (\textsc{Hin}) and Urdu (\textsc{Urd}) are mutually-intelligible yet sociolinguistically distinct registers of one Central Indo-Aryan language (usually termed Hindi--Urdu), spoken across the Indian subcontinent and by a majority in the northern part.~Gujarati (\textsc{Guj}) is a Central Indo-Aryan language and Marathi (\textsc{Mar}) is a Southern Indo-Aryan language, spoken in the western Indian states of Gujarat and Maharashtra, respectively.~Bengali (\textsc{Ben}), spoken in Bangladesh and the Indian state of West Bengal, and Odia (\textsc{Ori}), spoken in the Indian state of Odisha, are both Eastern Indo-Aryan languages. Finally, Tamil (\textsc{Tam}) and Malayalam (\textsc{Mal}) are both Dravidian languages spoken in the Indian states of Tamil Nadu and Kerala, respectively.~The Indo-Aryan and Dravidian language families are phylogenetically unrelated but have a long history of contact and cross-family bilingualism.

\subsection{Malayo-Polynesian}

Iban (\textsc{Iba}) is a Malayo-Polynesian language spoken by over 2 million people in Brunei as well as the Indonesian and Malaysian parts of the island of Borneo.~Iban has some use as a medium of education in the Malaysian state of Sarawak but does not possess official status.~It is related to Indonesian (\textsc{Ind}) and Malay (\textsc{Zsm}), which are the official languages of Indonesia and Malaysia, respectively.

\subsection{Sotho-Tswana}

Setswana (\textsc{Tsn}) is a Bantu language spoken by over 8 million people Botswana, South Africa, and Zimbabwe, where it is an official language.~Setswana also possesses minority language status in Namibia.~Two other languages of the Sotho-Tswana subgroup of Bantu are Sesotho (\textsc{Sot}, also known as ``Southern Sotho'') and Sepedi (\textsc{nso}, ``Northern Sotho'').~Sesotho is an official language of South Africa, Lesotho, and Zimbabwe, and Sepedi is an official language of South Africa.

\subsection{West Iberian}

Galician (\textsc{Glg}) is a Romance language spoken in Galicia, an administrative division of northwestern Spain bordering Portugal where it is the official language and spoken by over 2 million people.~It is closely related to Spanish (\textsc{Spa}) and Portuguese (\textsc{Por}), and all three are classified under the West Iberian subgroup of Romance languages. Galician and Portuguese split in the late Middle Ages (c.~15th century) and thus are the most closely related pair of the three.~Sociolinguistically, Galician speakers use Spanish in literary contexts and thus the two languages have a diglossic relationship.

\end{document}